\documentclass[preprint2]{aastex}
\begin{document}
\title{The Cepheids of Centaurus A (NGC 5128) and Implications for $H_0$}
\author{Daniel J. Majaess}
\affil{Saint Mary's University, Halifax, Nova Scotia, Canada}
\affil{The Abbey Ridge Observatory, Stillwater Lake, Nova Scotia, Canada}
\email{dmajaess@ap.smu.ca}

\begin{abstract}
A \textit{VI} Wesenheit and period-colour analysis based on new OGLE observations reaffirms Ferrarese et al. discovery of 5 Type II Cepheids in NGC 5128. The distance to that comparatively unreddened population is $d=3.8\pm 0.4 (\sigma_{\bar{x}})$ Mpc. The classical Cepheids in NGC 5128 are the most obscured in the extragalactic sample ($n=30$) surveyed, whereas groups of Cepheids tied to several SNe host galaxies feature negative reddenings. Adopting an anomalous extinction law for Cepheids in NGC 5128 owing to observations of SN 1986G ($R_V\simeq2.4$) is not favoured, granted SNe Ia may follow smaller $R_V$. The distances to classical Cepheids in NGC 5128 exhibit a dependence on colour and CCD chip, which may arise in part from photometric contamination. Applying a colour cut to mitigate contamination yields $d\simeq3.5$ Mpc ($\textit{V-I} \la1.3$), while the entire sample's mean is $d\simeq3.1$ Mpc. The distance was established via the latest \textit{VI} Galactic Wesenheit functions that include the 10 HST calibrators, which imply a shorter distance scale than Sandage et al.~(2004) by $\ga10$\% at $P\simeq25^d$. HST monitored classical Cepheids in NGC 5128, and the SNe hosts NGC 3021 \& NGC 1309, follow a shallower \textit{VI} Wesenheit slope than ground-based calibrations of the Milky Way, LMC, NGC 6822, SMC, and IC 1613. The discrepancy is unrelated to metallicity since the latter group share a common slope over a sizeable abundance baseline ($\alpha=-3.34\pm0.08(2\sigma)$, $\Delta$[Fe/H]$\simeq1$). A negligible distance offset between OGLE classical Cepheids and RR Lyrae variables in the LMC, SMC, and IC 1613 bolsters assertions that \textit{VI}-based Wesenheit functions are relatively insensitive to chemical abundance. In sum, a metallicity effect (\textit{VI}) is not the chief source of uncertainty associated with the Cepheid distance to NGC 5128 or the establishment of the Hubble constant, but rather it may be the admittedly challenging task of obtaining precise, commonly standardized, multiepoch, multiband, comparatively uncontaminated extragalactic Cepheid photometry.
\end{abstract} 

\section{Introduction}
\citet{fe07} discovered at least 51 classical Cepheids and 5 Type II Cepheid candidates in NGC 5128 (Centaurus A). The comprehensive survey provides an opportunity to ascertain the distance to NGC 5128 from population I \& II standard candles. That is particularly pertinent granted the classical Cepheid distance to NGC 5128 is inconsistent with independent indicators.  The discrepancy has been attributed to an anomalous extinction law and ambiguities surrounding the sensitivity of \textit{VI}-based Cepheid relations to chemical abundance.  Yet alternative rationale are favored in the present study.

Type II Cepheids continue to garner attention as a means of establishing the distances to globular clusters, the Galactic center, and galaxies \citep{ku03,ma09,ma09c,ma09d,ma10}.  Indeed, at least 21 Type II Cepheids were observed beyond the local group in M106 \citep{ma06,ma09c}.  The distance inferred to that galaxy from Type II Cepheids agrees with estimates established by masers and classical Cepheids \citep[$D_{TII}\simeq 7.3$ Mpc,][]{he99,ma06,ma09c}.  Discovering Type II Cepheids and RR Lyrae variables in galaxies hosting classical Cepheids offers an opportunity to constrain the effects of chemical composition on their luminosities and intrinsic colours \citep[][see also the historic precedent outlined in \citealt{ta08}]{fm96,ud01,ma09,ma09c,ma09d,ma10}.  However, the statistics must be conducive to the task, while the degeneracies posed by other uncertainties mitigated (e.g., photometric contamination via blending and crowding).  

In this study, additional evidence is presented to secure membership for 5 Type II Cepheid candidates observed by \citet{fe07} in NGC 5128 (\S \ref{stii}). In \S \ref{sdist} distances are computed for that galaxy's population of classical and Type II Cepheids, namely by employing: the latest \textit{VI} Galactic calibration which includes the new HST parallaxes for 10 nearby classical Cepheids; and a calibration inferred from recent \textit{VI} observations for 197 Type II Cepheids in the LMC (OGLE).  The associated uncertainties tied to the derived parameters are discussed, and pertain directly to the Cepheid distance scale and the establishment of $H_0$.  It is advocated that an anomalous extinction law (\S \ref{sdist}) and variations in chemical composition amongst Cepheids (\S \ref{smetallicity}) are unrelated to a significant disparity between the Cepheid distance to NGC 5128 and independent indicators.  The discrepancy may stem from the difficulties inherent to obtaining extragalactic Cepheid photometry (\S \ref{asources}).

\begin{figure}[!t]
\begin{center}
\includegraphics[width=6.5cm]{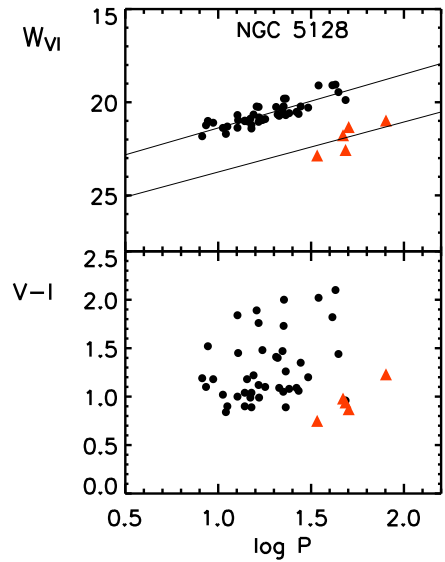}
\caption{\small{\textit{VI} Wesenheit and period-colour diagrams confirm \citealt{fe07} discovery of five Type II Cepheids in NGC 5128.  Type II and Classical Cepheids are indicated by red triangles and black dots accordingly, and are distinctly separated by $\simeq2$ magnitudes in Wesenheit space.  The Wesenheit magnitudes were evaluated as $W_{VI}=V-R_{VI} \times (V-I)$, where $R_{VI}=2.55$ is the canonical extinction law.  The slopes of the Wesenheit funcitons are variants of the LMC calibration \citep[][OGLE photometry]{ma09c}.  Long-period classical Cepheids in NGC 5128 exhibit a sizeable colour excess.}}
\label{fig1}
\end{center}
\end{figure}

\section{Type II Cepheids in NGC 5128}
\label{stii}
\citet{fe07} identified several potential Type II Cepheids in NGC 5128, with an emphasis placed on the following variables that exhibit Cepheid-like light curves: C43, C50, C52, C54, and C56.  However, the absence of a \textit{VI} calibrating dataset hampered efforts to secure the classification \citep[footnote 9,][]{fe07}.  The relevant data would be published a year later by the OGLE consortium who observed $197$ Type II LMC Cepheids in $V$ and $I$ \citep{so08}.   The candidates highlighted by \citet{fe07} may now be reassessed via \textit{VI} Wesenheit and period-colour diagrams (Fig.~\ref{fig1}).  There are drawbacks to applying only the aforementioned diagnostics to secure a Type II Cepheid designation \citep{ma09c}.  The former diagnostic is degenerate since variables of separate classes may overlap the Type II Cepheid Wesenheit relation, like semi-regulars \citep{so07,so08,so09b,pe09,ma09c}.  The latter diagnostic is problematic owing to differential reddening displacing a variable from the intrinsic or mean Type II Cepheid trend.  A strict adherence to the mean period-colour criterion led \citet{ma09c} to reduce their preliminary sample of $\gtrsim$100 extragalactic Type II Cepheids (excluding the LMC) by nearly $\simeq50$\%.  Additional diagnostics are needed which include period-amplitude and Fourier analyses of the light curves.  Yet RV Tau stars, which constitute the brightest subclass of Type II Cepheids \citep{sz10} and are therefore often detected in extragalactic surveys, exhibit somewhat chaotic and non-unique light curves that hamper efforts to secure a designation.  The matter is exacerbated since observations for Type II Cepheid candidates in remote galaxies are typically sparse and uncertain, particularly since the stars are often sampled fortuitously near the limiting magnitude of surveys seeking to discover brighter classical Cepheids.  

All the candidates highlighted by \citet{fe07} fall on the \textit{VI} Wesenheit relation characterizing Type II Cepheids (Fig.~\ref{fig1}).  The Wesenheit function is defined and discussed in \citet{vb68}, \citet{ma82}, \citet{op83,op88}, \citet{mf91,mf09}, and \citet{tu10}.  The relation is reddening-free and relatively insensitive to the width of the instability strip.  The population of Type II and classical Cepheids are distinctly separated by $\simeq2$ magnitudes in Wesenheit space.  BL Her, W Vir, and RV Tau stars do not follow the same linear \textit{VI} Wesenheit function \citep[][although see \citealt{mat06,mat09,fe10}]{so08}.  However, the linear relations displayed in Figure~\ref{fig1} merely identify and segregate the Cepheid populations \citep{ma09c}.  A separate relation that accounts for the reputed non-linearity of the \textit{VI} Type II Cepheid Wesenheit function is employed to establish distances \citep{ma09}.  

The \textit{VI} period-colour diagram demonstrates that the Type II Cepheid candidates exhibit apparent colours that are analogous to or somewhat bluer than their classical Cepheid counterparts (Fig.~\ref{fig1}).  That agrees with the trend noted for classical and Type II Cepheids in the LMC and M31 \citep[][photometry: \citealt{ud99,bo03}]{ma09c}.  Semi-regulars, by contrast, are typically redder than Cepheids.  The sparse sampling results in large uncertainties for the deduced mean magnitudes, periods, and hence classifications for the Type II Cepheid candidates.  The variables exhibit pulsation periods likely matching an RV Tau subclassification.  RV Tau stars may display alternating minima and maxima \citep[see the interesting discussion in][]{wo08}, however, that effect cannot be detected in the present data owing to the limited sampling and uncertainties (one cycle $\simeq 44^d$). 

\section{The Cepheid Distance to NGC 5128}
\label{sdist}
The distance to the Type II Cepheids in NGC 5128 may be ascertained via the \textit{VI} reddening-free relation established by \citet{ma09} from OGLE LMC calibrators \citep{ud99,so08}.  Likewise, the distance to the classical Cepheids may be computed using a \textit{VI} Galactic Cepheid calibration \citep{ma08}.  That calibration is based primarily on the efforts of fellow researchers who established classical Cepheids as members of Galactic open clusters \citep[e.g.,][]{sa58,mv75,tu92} or secured precise trigonometric parallaxes \citep[HST,][]{be07}.  The resulting \textit{mean} distance to the classical and Type II Cepheids in NGC 5128 is: $D_{TI}=3.06\pm  0.07 (\sigma_{\bar{x}} ) \pm 0.54 (\sigma )$ Mpc and $D_{TII}=3.8\pm 0.4 (\sigma_{\bar{x}}) \pm 0.8 (\sigma )$ Mpc.\footnote{$\sigma_{\bar{x}}$ and $\sigma$ are the internal standard error and standard deviation.  \citealt{fe07} error budget is provided in their Table~7.  Note that the Cepheid distances deviate as a function of colour and CCD chip by upwards of $\simeq0.4$ Mpc (\S \ref{asources}, Fig.~\ref{fig9}).} The classical Cepheid distance is essentially that determined by \citeauthor{fe07}, assuming the canonical extinction law ($D_{TI}\simeq3.1$ Mpc, see their Table 6).  The agreement is expected granted the Galactic classical Cepheid calibration yields a distance to the LMC of $\mu_0\simeq18.45$ \citep[\citealt{ma08}, \citealt{ma09d}, photometry:][]{ud99,se02,so08b}, which is comparable to the zero-point of the distance relation they employed.  \citet{fe07} consider and invariably adopt an anomalous extinction law for NGC 5128's Cepheids based on observations of supernova 1986G \citep[$R_V=2.4$,][]{ho87}, thereby increasing their estimate to $D_{TI}\simeq 3.4$ Mpc, which is the distance cited throughout the literature.  Yet recent observations indicate that SNe Ia may follow smaller $R_V$ than the canonical extinction law \citep{er06,wa06,go08,ng08}.  \citet{ri09b} cite a consensus value of $R_{V}\simeq2.5$ for SNe Ia which is consistent with that found by \citet{ho87} for the supernova in NGC 5128 (SN 1986G).  Adopting an anomalously low extinction law for Cepheids in NGC 5128 based solely on  observations of SN 1986G is not favoured. 

The \textit{mean} classical Cepheid distance to NGC 5128 disagrees with other indicators by $\simeq -20$\% \citep[][see also the NASA/IPAC Extragalactic Database (NED) master list of galaxy distances\footnote{http://nedwww.ipac.caltech.edu/level5/NED1D/intro.html} by \citealt{ms07}]{ha09}.  The distances cited above should be interpreted cautiously, irrespective of the aforementioned discrepancy.  The Type II Cepheid estimate \textit{presently} exhibits small statistics and large uncertainties, as expected.  The distances computed for the classical Cepheids exhibit a dependence on colour and CCD chip (\S \ref{asources}, Fig.~\ref{fig9}).  Additional concerns arise because that population is amongst the most obscured in the extragalactic sample (\S \ref{asources}). 

\begin{figure}[!t]
\begin{center}
\includegraphics[width=6.5cm]{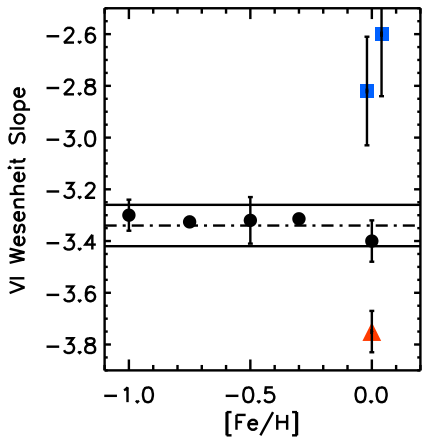}
\caption{\small{The slope of the \textit{VI} classical Cepheid Wesenheit relation is relatively insensitive to metallicity.  HST observations of classical Cepheids in NGC 5128, NGC 1309 and NGC 3021 (blue squares) follow a shallow slope by comparison to the latest ground-based observations of variables in the Milky Way, LMC, NGC 6822, SMC, and IC 1613 (black dots, $\alpha=-3.34\pm0.08 (2\sigma)$).  The slope of \citealt{sa04} Galactic calibration, based upon the best available data at the time of derivation and represented by the red triangle, disagrees with that inferred from the new HST parallaxes and (revised) cluster Cepheids  ($\alpha\simeq-3.4$).}}
\label{fig2}
\end{center}
\end{figure}

\section{Uncertainties associated with the Cepheid distance to NGC 5128}
\subsection{The (null) role of metallicity}
\label{smetallicity}
It has been argued that metal-rich classical Cepheids may exhibit a shallower (\& steeper) Wesenheit slope than metal-poor ones, thereby introducing a potential source of uncertainty into the present analysis since the chemical composition of the Cepheids in NGC 5128 is unknown.  However, a plot of the Wesenheit slopes inferred from ground-based observations of classical Cepheids in the Milky Way, LMC, NGC 6822, SMC, and IC 1613, demonstrates that the galaxies are characterized by a common \textit{VI} slope over a sizeable abundance baseline (Fig.~\ref{fig2}, $\alpha=-3.34\pm0.08(2\sigma)$ \& $\Delta$[Fe/H]$\simeq1$).  The slope of the \textit{VI} Wesenheit function is therefore insensitive to metallicity to within the uncertainties.  The contrasting interpretations and evidence presented by \citet{ta08} and \citet{ri09} should be considered.  

\begin{figure}[!t]
\begin{center}
\includegraphics[width=6.5cm]{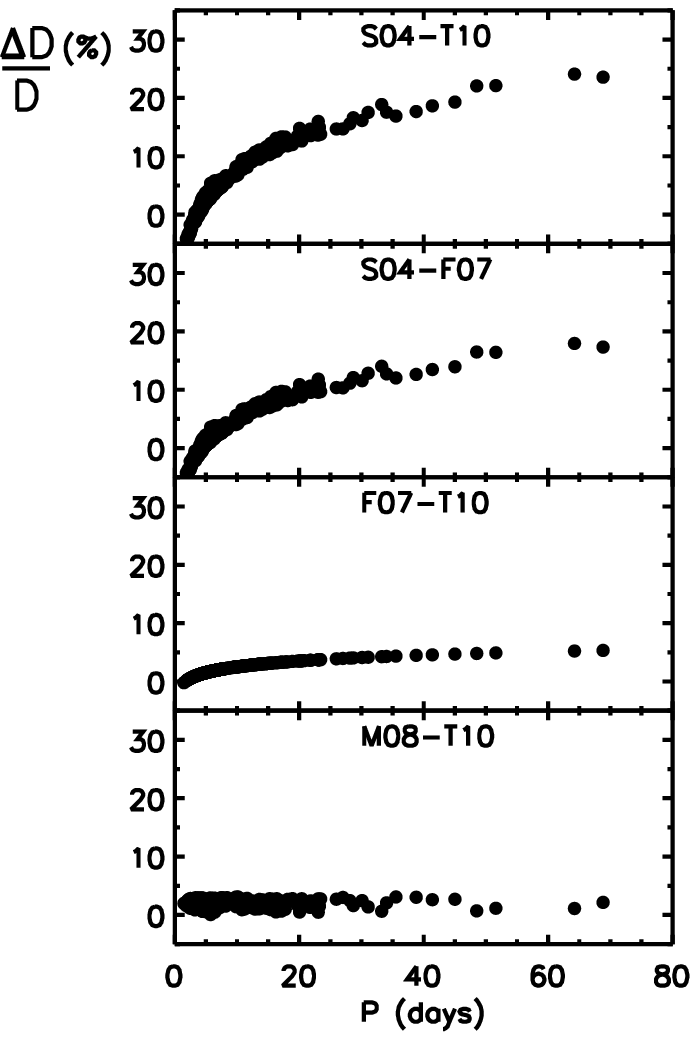}
\caption{\small{A comparison of the distances established to Galactic classical Cepheids via the \textit{VI} Wesenheit calibrations of \citealt{sa04} (S04), \citealt{fo07} (F07), \citealt{ma08} (M08), and \citealt{tu10} (T10).  The latter three calibrations include \citealt{be07} HST trigonometric parallaxes for 10 nearby classical Cepheids.  The \citealt{sa04} distance scale diverges from that of \citealt{fo07} and \citealt{tu10} by $\ga+10$\% at $P\simeq25^d$.}}
\label{fig3}
\end{center}
\end{figure}

The Galactic calibration employed to secure the distance to NGC 5128 and construct figure~\ref{fig2} is based in part on \citet{be07} HST parallaxes for 10 nearby classical Cepheids, which anchored the Milky Way calibration.  \citet{ta08} questioned the reliability of the HST parallaxes since the resulting period-$M_{V,I}$ relations inferred from that sample do not match their functions \citep{ta03,sa04},  which were constructed from the best available data at the time.  Their relations were derived prior to the publication of the HST parallaxes and the parameters for longer-period classical Cepheids tied to Galactic associations have since been revised \citep{tu10}, although continued work is needed to secure new calibrators and revise existing ones.\footnote{Facilitated by surveys initiated at the VISTA and OMM \citep{mi10,ar10}.}  The implied assertion that the HST parallaxes are awry is not supported by the results of \citet{tu10} or figure~\ref{fig2}.  A central conclusion of \citet{tu10} was that the classical Cepheid period-luminosity relation tied to the HST sample is in agreement with that inferred from cluster Cepheids. Moreover, the slope of the \textit{VI} Wesenheit function inferred from the HST parallaxes matches that of ground-based observations of classical Cepheids in the LMC, NGC 6822, SMC, and IC 1613 (Fig.~\ref{fig2}).  The \textit{VI} Galactic Wesenheit functions of \citet{fo07}, \citet{ma08}, and \citet{tu10} establish a distance scale which is $\ga10$\% nearer than \citet{sa04} at $P\simeq25^d$ (Fig.~\ref{fig3}).  The \textit{VI} Galactic Wesenheit calibration established by \citet{fo07}, partly on the basis of infrared surface brightness and interferometric Baade-Wesselink parallaxes, matches \citet{tu10} hybrid HST / cluster Cepheid based relation within $\la5$\% (Fig.~\ref{fig3}).   Lastly, regarding the construction of Fig.~\ref{fig2}, it is noted that the slope characterizing longer period Cepheids in IC 1613 is steeper than that describing the short period regime.  Moreover, the SMC exhibits a significant break in the \textit{VI} Wesenheit function \citep[see also][and references therein]{so10}.  The LMC displays a separate trend, and efforts continue to characterize the discrepancy and its source.   The reader is likewise referred to the research of \citeauthor{ng09}

\begin{figure*}[!t]
\begin{center}
\includegraphics[width=6.5cm]{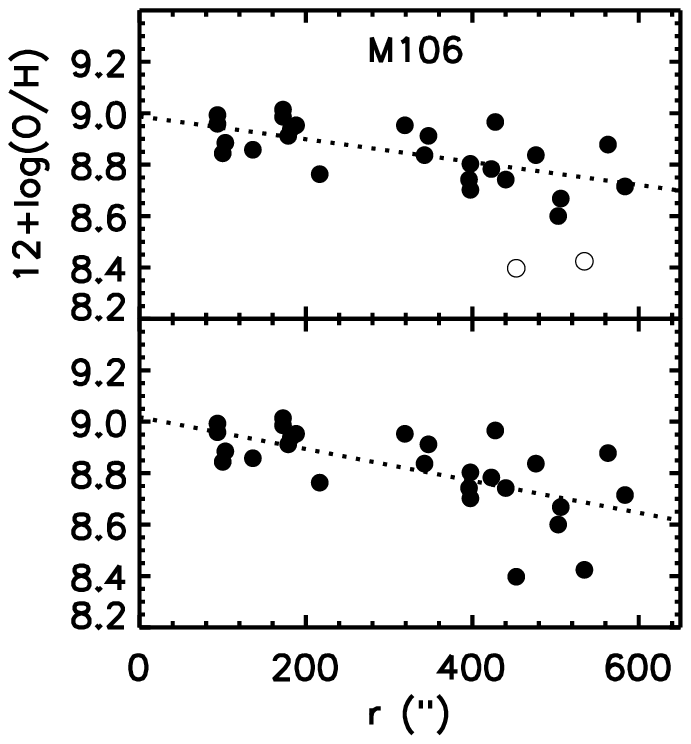}
\includegraphics[width=7cm]{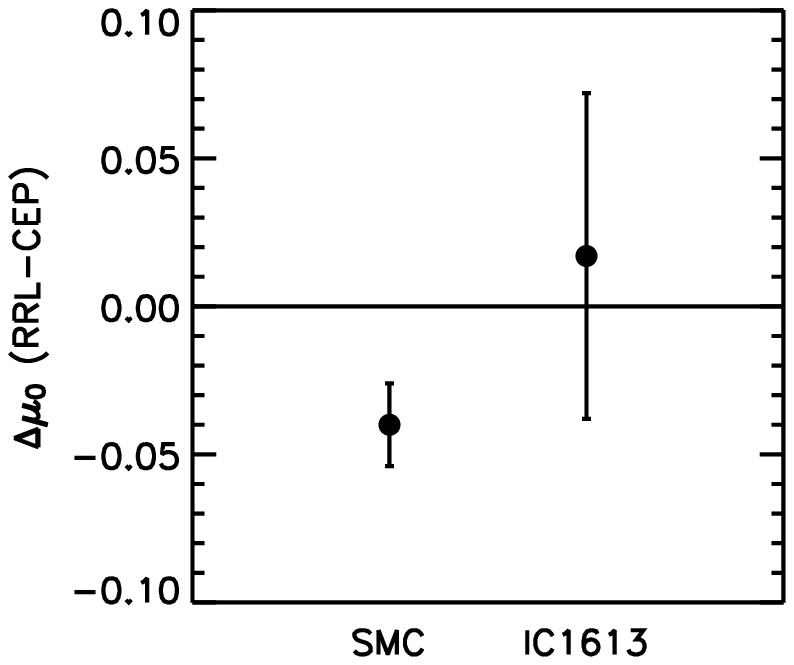}
\end{center}
\caption{\small{Left, \citealt{ri09} abundance gradient for M106 implies that initial estimates of the \textit{VI} classical Cepheid metallicity effect ($\gamma_i \simeq -0.3$ mag dex$^{-1}$) nearly double.  The exclusion of two datapoints implies an even larger value (left, top panel).  A sizeable metallicity effect contradicts evidence presented by a direct comparison of classical Cepheids, Type II Cepheids, and RR Lyrae variables at common zero-points.  Right, the distance offset between RR Lyrae variables and classical Cepheids in the SMC and IC 1613 is nearly negligible.  The base set of equations employed to compute the distances are OGLE \textit{VI} Wesenheit functions of LMC classical Cepheids and RR Lyrae variables.  The comparison is independent of zero-point and uncertainties tied to extinction corrections.  The results, in tandem with those of Fig.~\ref{fig2}, imply that the primary source of uncertainty tied to the Cepheid distance to NGC 5128 is unrelated to variations in chemical composition amongst Cepheids.}}
\label{fig7}
\end{figure*}

\citet{ke98}, \citet{ma06}, and \citet{sc09} suggest that the classical Cepheid \textit{VI} Wesenheit relation exhibits a zero-point dependence on metallicity \citep[see also the review of][]{ro05,ro08}, again introducing a potential source of uncertainty into the present analysis since the chemical composition of the Cepheids in NGC 5128 is unknown.  The aforementioned researchers endeavoured to ascertain the influence of chemical composition by examining the distance offset between classical Cepheids located in the central (metal-rich) and outer (metal-poor) regions of a particular galaxy (M101, M106, M33).  However, a degeneracy emerges (photometric contamination) since the stellar density and surface brightness often increase toward the central region.   \citet{mac01} noted that a substantial fraction of the difference in distance moduli between classical Cepheids occupying the inner and outer regions of M101 could arise from blending.  \citet{ma06} and \citet{sc09} employed criteria to mitigate the impact of photometric contamination so to enable an unbiased determination of the classical Cepheid metallicity effect from observations of classical Cepheids in M106 and M33, and the reader is encouraged to consider their evidence.  Yet the result inferred from variables in M106 was provided an alternative rationale by \citet{bo08} and \citet{ma09c}, who noted that the observed offset was too large to be attributed to variations in chemical composition.  Indeed, \citet{ri09} abundance gradient for M106 implies that initial estimates of the classical Cepheid metallicity effect ($\gamma_i \simeq -0.3$ mag dex$^{-1}$) nearly double \citep[Fig.~\ref{fig7}, or see Table 12 in][]{ri09}.  A comparably sizeable result is obtained when examining the offset between \citet{st98} distance to classical Cepheids occupying the inner region of M101 and \citet{ke96} distance to classical Cepheids in the outer region of that same galaxy, which sample metal-rich and metal-poor variables accordingly ($\gamma \simeq -0.5$ mag dex$^{-1}$, see also \citealt{ma09c}).  The results for M101 and M106 are larger than that cited for M33 ($\gamma \simeq -0.3$ mag dex$^{-1}$).  The results differ in yet another manner, namely that the slope of the \textit{VI} Wesenheit function inferred from classical Cepheids sampling the inner region of M106 differs from the outer region, while the classical Cepheids of M33 (inner \& outer) exhibit a comparable slope.  The discrepancies are manifold and the proposed metallicity effect is nonetheless too large.

The sizeable distance offset between the inner and outer regions of the galaxies arises from photometric contamination and other source(s).  Consider the following example, in tandem with the results of Fig.~\ref{fig2}, which compares the distances to classical Cepheids and RR Lyrae variables at a common zero-point (e.g., LMC, SMC, and IC 1613).  The \textit{VI} Wesenheit functions inferred from OGLE LMC classical Cepheids and RR Lyrae variables are adopted as the calibrating set \citep{ud99,so03}.  RR Lyrae variables likewise follow scatter reduced \textit{VI} Wesenheit functions \citep{kj97,so03,so09,di07,ma09d,ma10}.  The distance offset between classical Cepheids and RR Lyrae variables in the SMC as established via the OGLE LMC Wesenheit relations is: $\Delta \mu_0\simeq-0.04$ (Fig.~\ref{fig7}).  The distance offset between classical Cepheids and RR Lyrae variables in IC 1613 as established via the OGLE LMC Wesenheit relations is: $\Delta \mu_0\simeq+0.02$ (Fig.~\ref{fig7}).  The distances inferred from the  standard candles agree to within the uncertainties, despite the neglect of metallicity corrections for variable types sampling different temperature, radius, and density regimes. Hence the evidence does not support a sizeable metallicity effect.  The comparison between the variable types is independent of zero-point and uncertainties tied to extinction corrections. Admittedly, additional \textit{VI} observations of extragalactic RR Lyrae variables are desirable and the Wesenheit function characterizing that population as inferred from pulsation models, the Magellanic Clouds, and globular clusters are marginally discrepant \citep{kw01,di04,di07,so09}.  Further work is needed. 

In sum, metallicity does not significantly alter the \textit{VI} Wesenheit slope or zero-point \citep[Figs.~\ref{fig2} \& \ref{fig7}, see also][]{ud01,pi04,ma08,ma09,ma09c,bo08,ma09d,ma10}.  Therefore, concerns are allayed pertaining to chemical composition being a sizeable source of uncertainty tied to the Cepheid distance for NGC 5128, or the establishment of the Hubble constant.   By contrast, caution should be exhibited when employing \textit{BV} relations for Cepheids and RR Lyrae variables of differing abundance \citep[][and references therein]{ma09c}.  Caution is likewise urged when deriving a galaxy's distance and reddening via a multiwavelength approach which relies on Cepheid \textit{B}-band data.  

\begin{figure}[!t]
\begin{center}
\includegraphics[width=6cm]{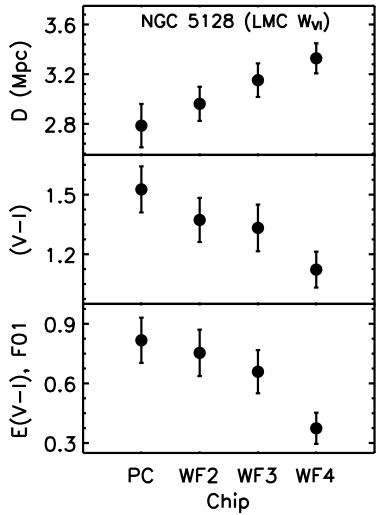}
\caption{\small{The distances of classical Cepheids in NGC 5128 exhibit a colour and CCD chip dependence, owing partly to photometric contamination.  A colour limited sample (mitigates contamination) yields $D\simeq3.5$ Mpc ($\textit{V-I} \la1.3$, \textit{see} Fig.~\ref{fig1}).  A \textit{VI} Wesenheit relation based on new OGLE\textit{III} observations was applied to infer the distance (LMC, $\mu_0=18.5$).}}
\label{fig9}
\end{center}
\end{figure}

\begin{figure*}[!t]
\begin{center}
\includegraphics[width=6.2cm]{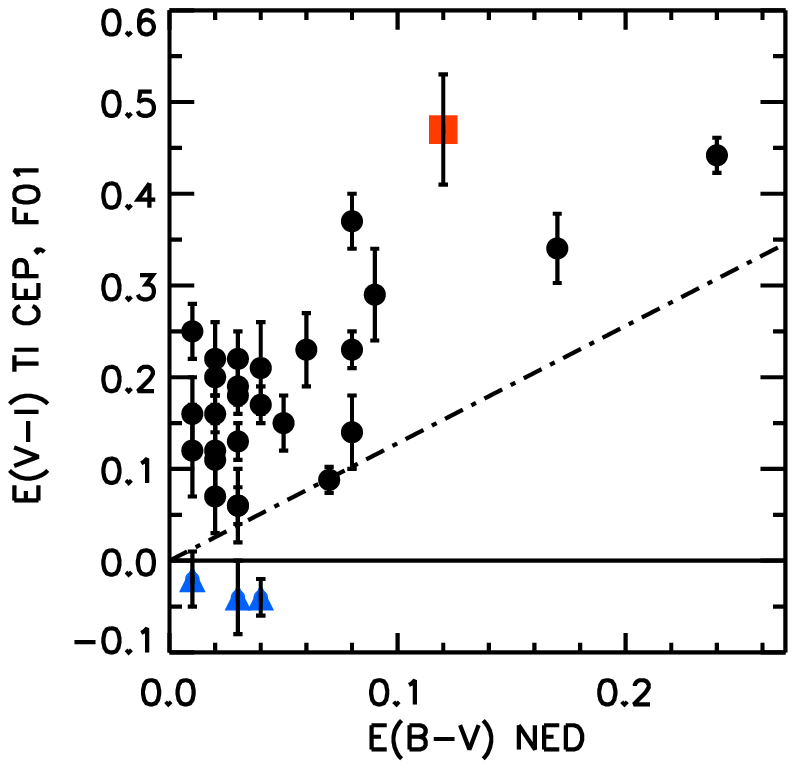}
\includegraphics[width=6.2cm]{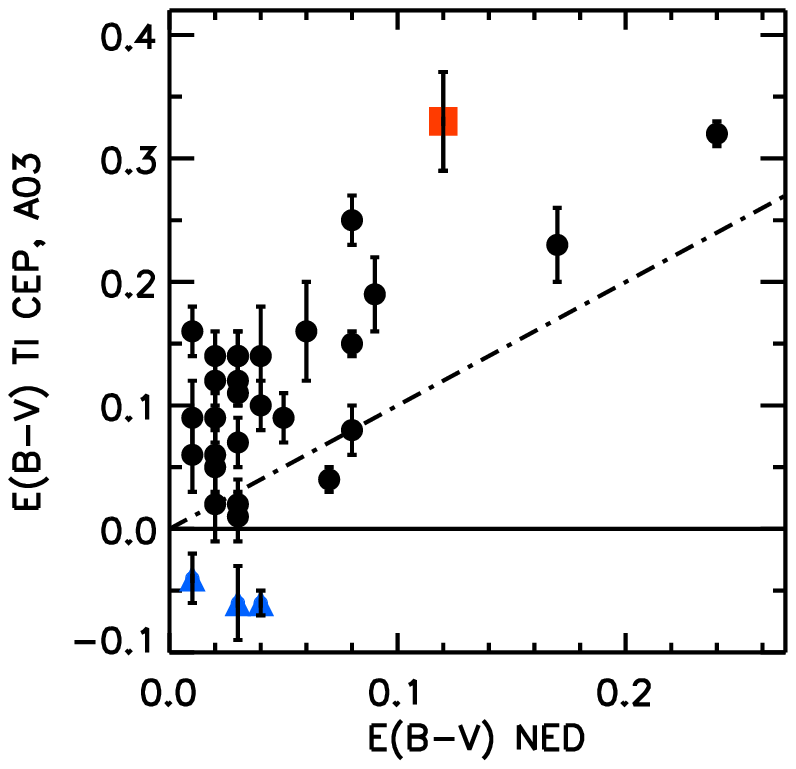} 
\includegraphics[width=6.2cm]{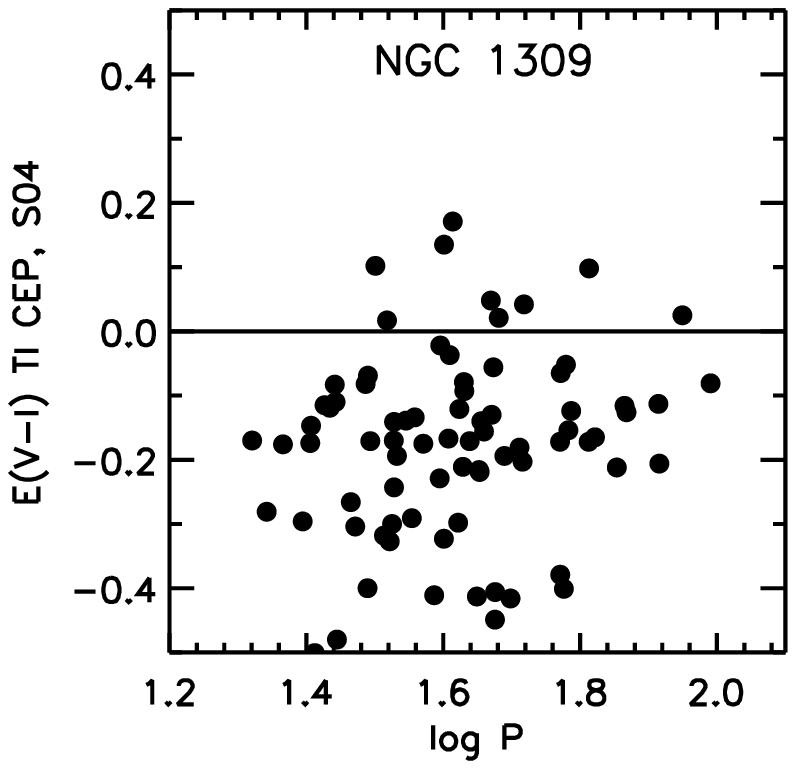} 
\includegraphics[width=6.2cm]{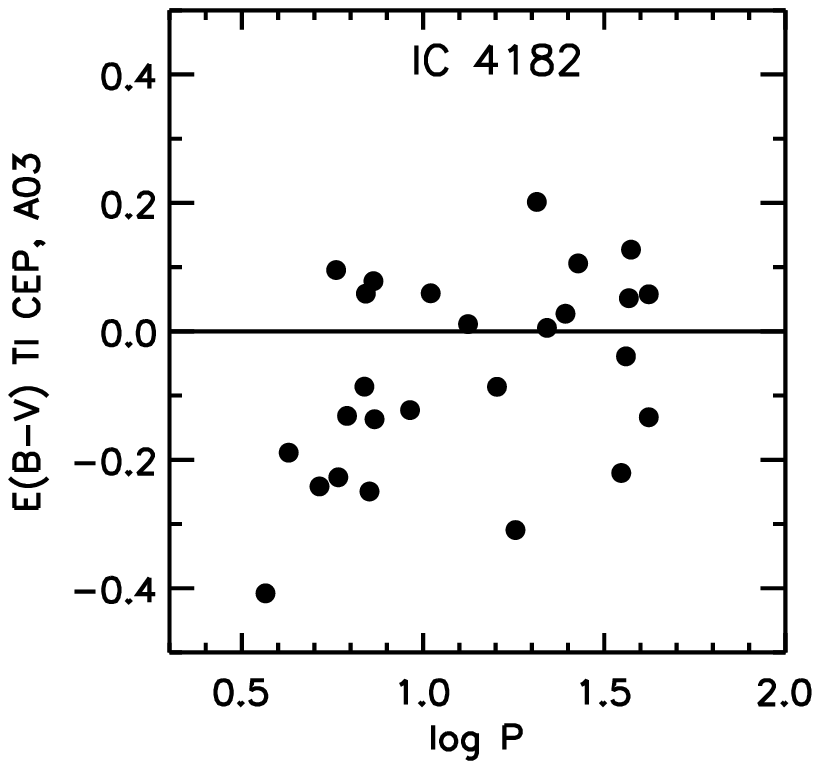} 
\caption{\small{Top panels, a comparison of the reddenings established for a sample of galaxies (including NGC 5128) from classical Cepheids and the NED extinction calculator.  E(V-I) is tabulated according to equations adopted by \citealt{fr01} (F01) and \citealt{sa04} (S04), whereas E(B-V) is computed following \citealt{ab03} (A03). Extragalactic classical Cepheid reddenings lie above the relation describing unity (dashed line).  The bulk of the data are offset $A_V\simeq0.^{m}3$ beyond the foreground estimate.  Classical Cepheids in NGC 5128 (red square) display a sizeable mean colour excess (see also Figs.~\ref{fig1},~\ref{fig9}).  By comparison, the Cepheid-SNe calibrating galaxies NGC 1309, NGC 3021, and IC 4182 (blue triangles) exhibit negative mean reddenings \citep[see also][]{sah06}.}}
\label{fig4}
\end{center}
\end{figure*}

\subsection{Extragalactic Cepheid photometry}
\label{asources}
Alternate sources that may explain the discrepancy between the Cepheid distance to NGC 5128 and independent indicators are now considered.  Of particular concern is the correlation between the computed distances to classical Cepheids in NGC 5128, their colours, and the sampling CCD (Fig.~\ref{fig9}).  The origin of the bias may be manifold.  

The excess reddening detected for a sizeable fraction of the classical Cepheids in NGC 5128 may be an indication of photometric contamination (Fig.~\ref{fig1}), which subsequently causes the affected stars to appear brighter and nearer \citep{su99,mo00,mo01}.  The most obscured Cepheids in the sample are issued the nearest distances.  Applying a colour cut as indicated by figure~\ref{fig1} yields $d\simeq3.5$ Mpc ($\textit{V-I} \la1.3$). The classical Cepheids of NGC 5128 exhibit the largest mean colour excess of the extragalactic sample examined (Fig.,~\ref{fig4}).   By contrast, negative mean reddenings were obtained for NGC 3021, NGC 1309, and IC 4182, galaxies which host classical Cepheids and SNe \citep[Fig.~\ref{fig4}, see also][]{sah06}.  The reddenings for the extragalactic sample (including NGC 5128) were established via the period-colour relations employed by \citet{fr01} and \citet{ab03}.   Applying \citet{sa04} period-$M_{V,I}$ relations would shift additional SNe-Cepheid calibrating galaxies into the negative absorption regime \citep[see also][]{sah06}.  That calibration yields a mean colour excess of $E_{V-I}\simeq-0.17$ for classical Cepheids in NGC 1309.  Period-colour relations do not account for the temperature dependence in the strip at a given period.  Consequently, reddenings computed for classical Cepheids on the hot edge of the strip will be overestimated, while reddenings computed for classical Cepheids on the cool edge of the strip will be underestimated.   The photometric errors inherent to extragalactic observations, in addition to internal differential reddening, exacerbate the perceived spread.  Period-color relations shall yield negative reddenings for Cepheids on the cool edge of the strip that are observed through negligible extinction, yet the mean for an entire sample of classical Cepheids should be null within the uncertainties owing to the even distribution of variables within the strip \citep{tu01}.   Suspicion should be cast upon photometry (\& the period-colour relations employed) which yield a mean extinction significantly less than the foreground estimate (Fig.~\ref{fig4}).  Also note that classical Cepheids observed in remote galaxies are preferentially the brightest (massive), and may be tied to star forming regions immersed in obscuring material (longer period classical Cepheids trace spiral arms: e.g., \citealt{ta70,be89,ma09,ma09b,ma10}). 

The presence of floating photometric zero-points is a concern owing to the difficulties inherent to achieving a common standardization, particularly across a range in colour and CCDs.  

The Wesenheit relations (LMC or MW) applied to infer the distance to NGC 5128 exhibit a steeper slope.  The Wesenheit slope describing classical Cepheids in NGC 5128 is $\alpha\simeq-2.9\pm0.3$ (Fig.~\ref{fig1}, sensitive to the sample and CCD chip chosen: Fig.~\ref{fig9}).  \citet{ri09} remarked that a sample of classical Cepheids in metal-rich galaxies hosting supernovae are likewise characterized by a shallow Wesenheit slope.   Photometric contamination, which may preferentially affect fainter short period Cepheids relative to brighter long period ones \citep[e.g., Fig.~17 in][]{ma06}, may bias the tilt of the inferred Wesenheit relation and could in part explain shallower slopes.  Applying an LMC or Galactic calibration to galaxies that exhibit vastly differing Wesenheit slopes shall introduce a global bias. Consider two galaxies sharing a common distance (e.g., the Leo I group) and spurious shallow Wesenheit slope, yet featuring variables of differing period distributions.  The galaxy containing the classical Cepheids characterized by a shorter period distribution shall be issued a nearer distance.  Moreover, an inhomogeneous period distribution across the CCD chips shall result in the propagation of artifical distance offsets across the detectors.

The effects described above may in sum conspire to produce figure~\ref{fig9}, and the discrepancy between the Cepheid distance to NGC 5128 and that established by independent means.  Admittedly, further work is needed to bolster the evidence.  

Lastly, the period-reddening function derived previously by the author \citep{ma09c} was not employed here because it has become apparent that the purely numerical method pursued to derive the relation was swayed by poor calibrating statistics toward the long period regime \citep{ma08}.  The Galactic classical Cepheid calibration employed by the author \citep{ma08} exhibits an absence of long period variables save $\ell$ Car, as perhaps too conservative a philosophy was imposed requiring cluster Cepheids enlisted in the calibration be secured via radial velocities or proper motions.  A bias is introduced since $\ell$ Car lies well toward the red edge of the instability strip \citep[see][]{tu10}.  The author shall revisit the \textit{VI} period-reddening formalism and subject elsewhere, an analysis that shall be facilitated by the recent establishment of spectroscopic reddenings for a sizeable sample of Galactic classical Cepheids \citep{ko08}. 

\section{Summary \& Future Research}
The properties of classical and Type II Cepheids in NGC 5128 are reinvestigated by employing calibrations featuring the latest OGLE \& HST data.  Sources beyond an anomalous extinction law or variations in chemical composition amongst Cepheids are proposed to rationalize the significant discrepancy between the Cepheid distance to NGC 5128 and other indicators.

Five Type II Cepheid candidates discovered by \citet{fe07} in NGC 5128 exhibit \textit{VI} Wesenheit magnitudes and colours that are consistent with the proposed designation (Fig.~\ref{fig1}).  The pulsation periods could imply an RV Tau subclassification.  RV Tau stars may exhibit alternating minima and maxima  \citep[see][]{wo08}, however, the presence of that effect cannot be ascertained because the observational baseline is only one cycle ($\simeq44^d$).  The Type II Cepheids are observed through marginal extinction, in contrast to their classical Cepheids counterparts (Figs.~\ref{fig1},~\ref{fig4}).

The \textit{mean} distance to NGC 5128's population of Type II and classical Cepheids is: $D_{TII}=3.8\pm 0.4 (\sigma_{\bar{x}}) \pm 0.8 (\sigma )$ Mpc and $D_{TI}=3.06\pm  0.07 (\sigma_{\bar{x}} ) \pm 0.54 (\sigma )$ Mpc.  The latter estimate is essentially that obtained by \citet{fe07} while employing the canonical extinction law \citep[$R_V\simeq3.3$, Table~6 in][]{fe07}.  Adopting an anomalous extinction law for classical Cepheids in NGC 5128 owing to observations of SN 1982G ($R_V\simeq2.4$) is not favoured.  SNe Ia may follow smaller $R_V$ than the canonical value \citep{er06,wa06,go08,ng08}.  Small statistics \textit{presently} dominate the uncertainty of the Type II Cepheid distance to NGC 5128, mitigating the estimate's importance. The \textit{mean} classical Cepheid distance to NGC 5128 ($D_{TI}\simeq3.1$ Mpc) disagrees with other indicators by $\ge-20$\% \citep{ha09}.  

The distance to the classical Cepheids in NGC 5128 was determined by applying the latest \textit{VI} Galactic Wesenheit calibrations which utilize \citet{be07} new HST trigonometric parallaxes for 10 nearby classical Cepheids \citep[e.g.,][]{fo07,ma08,tu10}.  The distance scale implied by the \textit{VI} Galactic calibrations of \citet{fo07} and \citet{tu10} are $\ga10$\% less than that advocated by \citet{sa04} at $P\simeq25^d$ (Fig.~\ref{fig3}).  The \citet{ta03} and \citet{sa04} relations were constructed prior to the publication of \citet{be07} HST parallaxes for 10 nearby classical Cepheids, which anchored the Milky Way calibration. \citet{sa04} relation forms the basis for \citet{san06} estimate of $H_0\simeq62$ km s$^{-1}$ Mpc$^{-1}$.  That estimate is smaller than the value espoused by \citet{fr01} or \citeauthor{ri09b}, and the discrepancy hampers efforts to constrain cosmological models \citep[][Table~1]{ri09b}.   The difference amongst the \textit{VI} Galactic calibrations cited above may explain the bulk of the disagreement between the estimates of $H_0$, however, that conclusion is somewhat presumptuous.  Redetermining $H_0$ to compliment the aforementioned estimates is desirable, but requires a scrupulous inspection of all the archived data while considering advances in the field and the assertions summarized here, an effort which may be pursued elsewhere.   For example, added weight shall be given to calibrating galaxies which exhibit: a population of classical Cepheids that are characterized by a \textit{VI} Wesenheit slope near $\alpha\simeq-3.34\pm0.08(2\sigma)$ (Fig.~\ref{fig2}), where the slope is not imposed upon the data unless reaffirmed by a least-squares fit; Cepheids that are issued consistent moduli across the CCD chips; Cepheids that are observed through low obscuration, yet marginally greater than the foreground extinction estimate (Fig.~\ref{fig4}); Cepheids that are sampled in low density and low surface brightness environments so to mitigate photometric contamination, etc.  Period-colour relations and the slope of the \textit{VI} Wesenheit function may be employed to screen photometry and assess quality. 

The \textit{VI} Wesenheit functions describing classical Cepheids in NGC 5128, and the SNe hosts NGC 3021 \& NGC 1309, exhibit a shallower slope than  calibrations of the Milky Way, LMC, NGC 6822, SMC, and IC 1613 (Fig.~\ref{fig2}).  The discrepancy is not tied to variations in metallicity since ground-based observations of classical Cepheids in the Milky Way, LMC, NGC 6822, SMC, and IC 1613 feature comparable \textit{VI} Wesenheit slopes over a sizeable abundance baseline (Fig.~\ref{fig2}, $\alpha=-3.34\pm0.08(2\sigma)$, $\Delta$[Fe/H]$\simeq1$).  The aforementioned galaxies exhibit the most precise photometry of all the Cepheid data inspected (Figs.~\ref{fig4}).  The distances computed for classical Cepheids in NGC 5128 display a dependence on colour and CCD chip, which is likely attributable in part to photometric contamination (Figs.~\ref{fig1} \& \ref{fig9}, see text).  Applying a colour cut yields $d\simeq3.5$ Mpc ($\textit{V-I} \la1.3$, see Fig.~\ref{fig1}).  The classical Cepheids otherwise exhibit the largest mean colour excess of the extragalactic sample examined (Figs.~\ref{fig1},~\ref{fig4}).  By contrast, and perhaps disconcertingly, Cepheids tied to several galaxies hosting SNe feature negative (or near negligible) mean reddenings \citep[Fig.~\ref{fig4}, see also][]{sah06}.  The extragalactic classical Cepheid sample displays a mean $A_V\simeq0.^{m}3$ offset beyond the foreground extinction estimate inferred from dust maps (Fig.~\ref{fig4}).  Fig.~\ref{fig4} reaffirms that reddenings inferred from foreground dust extinction maps for distant galaxies are likely underestimated. 

A zero-point metallicity correction is not the chief source of uncertainty tied to the \textit{VI}-based Cepheid distance for NGC 5128, or the establishment of the Hubble constant (\textit{VI} photometry).  \citet{ri09} abundance gradient for M106 implies that initial estimates of the classical Cepheid metallicity effect nearly double (Fig.~\ref{fig7}).  That value is too large, and contradicts a direct comparison of OGLE classical Cepheids and RR Lyrae variables in the Magellanic Clouds and IC 1613 which exhibit a negligible distance offset ($\Delta \mu_0\simeq+0.01\pm0.06$, Fig.~\ref{fig7}).  Moreover, the metallicity effect cited in the literature and inferred from observations of M33, M101, and M106 is discrepant \textit{vis \`a vis} both the zero-point and slope dependencies.  In sum, the evidence indicates that the slope and zero-point of the classical Cepheid \textit{VI} Wesenheit function are largely insensitive to variations in chemical abundance \citep[Figs.~\ref{fig2} \& \ref{fig7}, see also][]{ud01,pi04,ma08,ma09,ma09c,bo08,ma09d,ma10}.   A primary source of uncertainty tied to the Cepheid distance to NGC 5128, and that which hampers efforts to constrain cosmological models, may be the admittedly challenging task of obtaining precise, commonly standardized, multiepoch, multiband, comparatively uncontaminated extragalactic Cepheid photometry. 

\subsection*{acknowledgements}
\scriptsize{The author is grateful to fellow researchers L. Ferrarese, A. Riess, A. Saha, G. Pietrzy{\'n}ski, F. Benedict, A. Dolphin, I. Soszy{\'n}ski \& the OGLE team, whose comprehensive surveys were the foundation of the study, the AAVSO (M. Saladyga, A. Henden), CDS, arXiv, NASA ADS, NED (I. Steer), and the RASC.}

\end{document}